# A tail dependence-based MST and their topological indicators in modelling systemic risk in the European insurance sector

Anna Denkowska[1] and Stanisław Wanat[2]

**Abstract**

In the present work we analyse the dynamics of indirect connections between insurance companies that result from market price channels. In our analysis we assume that the stock quotations of insurance companies reflect market sentiments which constitute a very important systemic risk factor. Interlinkages between insurers and their dynamics have a direct impact on systemic risk contagion in the insurance sector. We propose herein a new hybrid approach to the analysis of interlinkages dynamics based on combining the copula-DCC-GARCH model and Minimum Spanning Trees (MST). Using the copula-DCC-GARCH model we determine the tail dependence coefficients. Then, for each analysed period we construct MST based on these coefficients. The dynamics is analysed by means of time series of selected topological indicators of the MSTs in the years 2005-2019. Our empirical results show the usefulness of the proposed approach to the analysis of systemic risk in the insurance sector. The times series obtained from the proposed hybrid approach reflect the phenomena occurring on the market. The analysed MST topological indicators can be considered as systemic risk predictors.

1. Introduction.

Currently, despite many studies that use different methodological and empirical approaches to identify and analyze systemic risk in the insurance sector, there is still no consistent theory to monitor it effectively. Ideal methods that could be used for this purpose should support or be associated with the essential elements of macroprudential policy and surveillance (MPS) by providing information on the build-up of system-wide vulnerabilities in time and cross-section, with an acceptable level of accuracy for both the forecast of the occurrence of a systemic event and its financial effects. The subject of this article fits into the current of research focusing on the search for such a method. We focus on the structure of interlinkages between insurance companies, which plays a key role in the spread of systemic risk in this sector.

[1] Department of Mathematics, Cracow University of Economics, Kraków, Poland.
   E-mail: anna.denkowska@uek.krakow.pl. ORCID ID: https://orcid.org/0000-0003-4308-8180
[2] Department of Mathematics, Cracow University of Economics, Kraków, Poland.
   E-mail: wanats@uek.krakow.pl. ORCID ID: https://orcid.org/0000-0001-6225-4207

The article is a response to the clue and task left in the work of Alves et al. (2015), which appeared in the European System Risk Board. This work contains an analysis of the network of 29 largest European insurance groups and their financial contractors. The authors note that insurance companies have direct exposures to other insurers, banks and other financial institutions through debt, equity and other financial instruments. These exposures can cause direct infection and thus spread of systemic risk. The work cited above focuses on direct connections between EU insurers and banks. At the same time, the authors emphasize that their research does not include an analysis of linkages between insurers under reinsurance contracts, indirect connections via market price channels and information channels, nor an analysis of banks' exposure to insurers.

In the present article, we focus on the problem of indirect links between insurance companies that result from market price channels. More specifically, we examine the dynamics of these relationships. In our analysis, we assume that stock market quotations of insurance companies reflect market sentiment, which is a very important systemic risk factor. It is well known that risk infection is always accompanied by negative market sentiment leading to customer panic in the financial industry. The results will create a vicious circle of risk and emotion. Thus, market sentiment is commonly used to forecast changes in the financial market and can be used as a systemic risk barometer (Kou et al., 2019).

Relationships between insurers and their dynamics have a direct impact on propagation of systemic risk in the insurance sector. In our work, we propose a new hybrid approach to analyzing the dynamics of interconnections, based on combining the copula-DCC-GARCH and minimum spanning trees (MST). Using the copula-DCC-GARCH model, we determine tail dependence coefficients. Then, based on these coefficients for each analyzed period, we determine the "distance" matrix between insurance companies using the Mantegna metric (Mantegna and Stanley, 1999) and construct minimum spanning trees. We analyze the dynamics using selected topological indicators for the MST obtained.

The main purpose of the work and contribution to the literature is:
1) to check whether the time series of topological indicators of the network of connections between insurance companies obtained using the proposed hybrid approach reflect the situation on the financial market and whether they can be used as predictors of systemic risk in the insurance sector.

2) an empirical analysis of 38 European insurance institutions selected from the top 50 insurance companies in Europe. We indicate which of the largest companies not on the G-SIIs list are of great importance in the context of SR.

3) an analysis of the situation in the insurance sector in the context of SR, taking into account the latest political and economic situation in Europe, distinguishing four market states: the normal state, the state related to the Subprime Mortgage Crisis, the state related to the immigration crisis in Europe, and the state related to the crises in France and Italy.

4) an analysis of the contribution to the SR of the insurance sector.

The rest of the article is organized as follows. The second section reviews the subject literature devoted to systemic risk in the insurance sector. The third section presents the methodology and the empirical strategy used in the paper, the fourth one contains the data and a discussion of the results obtained, while the fifth and last one presents the conclusions.

2. Systemic risk in the insurance sector.

For over a decade, scientists have been trying to effectively define, study and measure the phenomenon of systemic risk, which in the era of globalization of economics is one of the most important concepts in the prediction of economic phenomena. Most scholars base their definition of uncertainty and risk on Knight (1921, p. 233), Tversky and Kahneman (1992), Camerer and Weber (1992) and Zweifel and Eisen (2012, p. 1). In the work Eling and Pankoke (2016) 43 definitions are given, which indicate a three-stage course of the phenomenon: causes, event and effects for the real economy. One of the latest approaches is the concept of systemic risk proposed by De Bandt and Hartmann (2000), in which a distinction is made between the risk of shocks based on their second-round effects (it focuses not on the institutions affected by shock, but on the consequences on linked institution). In addition, Harrington (2009) distinguishes between systemic risk and the risk of typical shocks. According to him, only the risk of an event associated with 'cross-contagious infection' (p. 802) should be considered systemic. Many researchers analyze the problem of SR in the context of the failure of a significant part of the financial sector and reduction of credit availability, e.g. Acharya et al.

(2011). Adrian and Brunnermeier (2011) investigate the negative impact on credit supply, Bach and Nyuyen (2012), Rodríguez-Moreno and Peña (2013) financial system failure, Baur et. al. (2003), Chen et al. (2013b), Cummins and Weiss (2011, 2013), Weiß and Mühnickel (2014) the negative impact on the real economy, Baluch et al. (2011) the chain reaction of financial difficulties, Chen et al. (2013a), Huang et al. (2009) many simultaneously defaulted pledges by large financial institutions, IAIS (2009), Jobst (2014), Radice (2010) the disruption of the flow of financial services, the negative impact on the real economy and impairment of all or part of the financial system, Klein (2011) studies the market in the context of financial system instability, idiosyncratic events and infection, Kress (2011) studies infection, Rodríguez-Moreno and Peña (2013) malfunctioning in the financial system and the negative impact on the real economy. In recent years, quantitative analysis of systemic risk using the described approaches has been carried out by, among others, Hautsch et al. (2015), Giglio et al. (2016), Benoit et al. (2017), Jajuga al. (2017), Bégin et al. (2017), Jurkowska (2018).

The various concepts of systemic risk analysis presented above have inspired the creation of a number of different methods for measuring it. In the literature of the subject, several dozen measures can be indicated, which can be determined using mathematical, statistical, econometric, network modeling and predictive analysis tools (in particular, multidimensional statistical analysis, including methods of learning with and without supervision). A review of systemic risk measures in use can be found e.g in the following articles: Bisias et al. (2012), Giglio et al. (2016), Di Cesare and Rogantini (2018).

It is worth noting that while there is quite extensive a literature on the subject of systemic risk analysis in the banking sector, the insurance sector has been analyzed to a distinctly smaller extent. The reason for this was the belief that the group taking over, dispersing and redistributing the financial effects of risk does not generate a systemic threat.

However, after the financial crisis in 2007-2009 and the European public debt crisis in 2010-2012, a significant increase in interest in systemic risk in the insurance sector can be seen. Before the crisis, there was a clear belief among researchers that this sector is systemically insignificant. However, in the literature that emerged as a result of the crisis, although previous conviction was maintained in many studies, there appeared articles indicating the possibility of the insurance sector creating systemic risk. Examples include works in which the authors believe that insurance companies have become an unavoidable source of systemic risk (e.g. Billio et al. 2012, Weiß and Mühlnickel 2014) and in which they claim insurance companies to be systematically significant, but only due to their non-traditional (banking) activities (e.g. Baluch et al. 2011, Bednarczyk 2013, Cummins and Weiss 2014, Czerwińska 2014) and the overall systemic importance of the insurance sector as a whole is still subdued to the banking sector (e.g. Chen et al., 2013). In turn, Bierth et al. (2015) after examining a very large sample of insurers in the long term, believe that the contribution of the insurance sector to systemic risk is relatively small, however, they claim that it peaked during the financial crisis in the period from 2007 to 2008. They also indicate that significant factors affecting the insurers' exposure to systemic risk are strong linkages between large insurance companies, leverage, losses and liquidity (the four L's: linkages, leverage, losses, liquidity). On the other hand, there are also studies (Harrington 2009) and (Bell and Keller 2009) claiming a complete lack of evidence for the systemic importance of the insurance industry.

After the aforementioned crises, supervisory authorities also began to pay more attention to the problem of systemic risk in the insurance industry. The Financial Stability Board (FSB), in consultation with the International Association of Insurance Supervisors (IAIS), identified nine global systemically important insurers (G-SIIs) based on the assessment methodology developed by IAIS, which includes the following five elements: non-insurance activity of the insurer (45%), assessment of the degree of direct and indirect links of institutions within the financial system (40%), range of global activity (5%), the size of the insurance institution (5%), product substitutability (5%).

Since the publication of this methodology and the G-SII list, questions have been raised about the appropriateness and effectiveness of the proposed framework by both the insurance sector and academia. The ongoing discussion in the literature to date tend to show that some indicators in the IAIS assessment methodology may not be able to explain the insurers' contribution to systemic risk (Weiß and Mühlnickel 2014, Bierth et al. 2015). Looking at the solutions adopted from the insurance industry, one can point to the example of MetLife, which was constantly struggling to remove the label of a "systemically important institution" and obtained a favorable ruling in the US District Court in March 2016 (Wall Street Journal 2016). Following the success of MetLife, the AIG SIFI label was withdrawn by FSOC in September 2017, and Prudential Financial dumped its brand SIFI in October 2018.

To sum up, current literature and real events show that systemic risk in the insurance sector is still a challenge waiting for precise methodological solutions. After 2014, we observe an increased involvement of scientists in the qualitative and quantitative analysis of this issue. Our paper is one of the few quantitative studies on systemic risk in the European and global insurance sector. Although SR in in the financial sector is analyzed by: Bierth et al. (2015), Mühlnickel and Weiß (2015), Kanno (2016), Giglio et al. (2016), Adrian and Brunnermeier (2016), Koijen (2016), Brownlees (2017), Kaserer (2018) and risk infection is studied by Hautsch et al. (2015), Härdle et al. (2016), Fan et al. (2018), nevertheless, none of these approaches is a hybrid approach in which the possibility of combining different measures would be analyzed on such a scale as proposed in our project.

3. Methodology.

We carry out the analysis of the dynamics of interconnections between insurance companies using a new hybrid approach based on the combination of the copula-DCC-GARCH model and minimum spanning trees (MST). The construction of minimum spanning trees based on the dependencies in the tails plays a key role in it. To this end, using two-dimensional copula-DCC-GARCH models for each studied period $t$, $(t = 1, ..., T)$ and each pair of rates of return $r_{i,t}, r_{j,t}$, $(i, j = 1, ..., k, j > i)$ we estimate the bivariate joint distributions:

$$F_t(r_{i,t}, r_{j,t}) = C_{ij,t}\left(F_{i,t}(r_{j,t}), F_{j,t}(r_{i,t})\right), \qquad (1)$$

where $C_{ij,t}$ denotes the copula, while $F_t$ and $F_{i,t}, F_{j,t}$, respectively, are the joint cumulative distribution function and the cumulative distribution functions (*cdf*) of the marginal distributions at time $t$. In turn, making use of the copulas $C_{ij,t}$ we estimate the pairwise lower tail dependence of the returns $r_{i,t}, r_{j,t}$:

$$\lambda_t^L(i,j) = \lim_{q \to 0^+} \frac{C_{ij,t}(q,q)}{q}. \qquad (2)$$

Then, for each period *t*, we determine the "distance" matrix between insurance companies using the metric (Mantegna, Stanley, 1999):

$$d_t(i,j) = \sqrt{2(1 - \lambda_t^L(i,j))} \qquad (3)$$

and using the Kruskal algorithm (Mantegna and Stanley, 1999), we construct minimum spanning trees $MST_t$ with $k$ vertices and $k - 1$ edges.

Based on the trees thus obtained $MST_t$ $(t = 1 ... T)$ we determine the time series of the following topological network indicators:

- Average Path Length - APL,
- Maximum Degree - Max.Deg,
- The parameters α of the vertex degree distribution required to follow a power law,
- Network Diameter - D,
- Rich Club Effect - RCE,
- Assortativity,
- Betweenness Centrality - BC,
- Vertex Strength (Centrality),

- Vertex Degree,
- Closeness Centrality.

It should be mentioned that in the literature the minimum spanning trees that evolve in time are also monitored by many other topological network indicators such as the Eigenvector Centrality (Tang et al. 2018), MOL (Mean Occupation Layer) (Onnela et al. 2002, 2003), Normalized Tree Length (Onnela et al. 2003), Tree Half-life (Onnela et al. 2003); Survival Ratio of the edges (Onnela et al. 2002, Sensoy and Tabak 2014); and Agglomerative Coefficient (Matesanz and Ortega 2015).

In the next stage of research, we determine a time series for the deltaCoVaR measure for each insurer. It brings along an information about the insurer's contribution to the systemic risk in the insurance sector. For this purpose, we also use two-dimensional copula-DCC-GARCH models and the empirical strategy presented in the articles: Denkowska and Wanat (2019), Wanat and Denkowska (2018a, 2018b, 2019). We assume that the European insurance sector is represented by the STOXX 600 Europe Insurance index. We compare the time series of deltaCoVaR measures obtained in this way with the time series of topological indicators of the $MST_t$ from the point of view of the possibility of using the latter as systemic risk predictors in the insurance sector.

The tail dependence coefficients ($\lambda_t^L(i,j)$) and the deltaCoVaR measure, which are key to the empirical strategy presented above, are determined using two-dimensional copula-DCC-GARCH models. In the k-dimensional case in the DCC-GARCH model, the returns vector distribution $r_t = (r_{1,t}, \ldots, r_{k,t})$ which is conditional with respect to the set $\Omega_{t-1}$ of information available up to the moment $t-1$ is modelled using the conditional copulas proposed by Patton (2006). It takes the following form:

$$r_{1,t}|\Omega_{t-1} \sim F_{1,t}(\cdot\,|\Omega_{t-1}), \ldots, r_{k,t}|\Omega_{t-1} \sim F_{k,t}(\cdot\,|\Omega_{t-1})$$
$$r_t|\Omega_{t-1} \sim F_t(\cdot\,|\Omega_{t-1})$$
$$F_t(r_t|\Omega_{t-1}) = C_t\left(F_{1,t}(r_{1,t}|\Omega_{t-1}), \ldots, F_{k,t}(r_{k,t}|\Omega_{t-1})\right),$$

Where $C_t$ denotes the copula, while $F_t$ and $F_{i,t}$ – respectively, the multivariate distribution and the marginal distributions at the moment *t*. In general, one-dimensional returns can be modeled using different specifications of the average model and different specifications of the variance model (e.g. sGARCH, fGARCH, eGARCH, gjrGARCH, apARCH, iGARCH, csGARCH). In our study the following ARMA process was used for all the series of returns:

$$r_{i,t} = \mu_{i,t} + y_{i,t},$$

$$\mu_{i,t} = E(r_{i,t}|\Omega_{t-1}),$$

$$\mu_{i,t} = \mu_{i,0} + \sum_{j=1}^{p_i} \varphi_{ij} r_{i,t-j} + \sum_{j=1}^{q_i} \theta_{ij} y_{i,t-j},$$

$$y_{i,t} = \sqrt{h_{i,t}} \varepsilon_{i,t},$$

and the standard GARCH (sGARCH) model for the variance:

$$h_{i,t} = Var(r_{i,t}|\Omega_{t-1}), \qquad h_{i,t} = \omega_i + \sum_{j=1}^{p_i} \alpha_{ij} y_{i,t-j}^2 + \sum_{j=1}^{q_i} \beta_{ij} h_{i,t-j}$$

where $\varepsilon_{i,t} = \frac{y_{i,t}}{\sqrt{h_{i,t}}}$ are identically distributed independent random variables (in the empirical analysis we considered the following distributions: normal, skew normal, t-Student, skew t-Student and GED). To describe the dependences between the rates of returns we used Student t-copulas, whose parameters were the conditional correlations $R_t$ obtained from the $DCC(m,n)$ model:

$$H_t = D_t R_t D_t,$$

$$D_t = diag(\sqrt{h_{1,t}}, \dots, \sqrt{h_{k,t}}),$$

$$R_t = (diag(Q_t))^{-\frac{1}{2}} Q_t (diag(Q_t))^{-\frac{1}{2}}$$

$$Q_t = (1 - \sum_{j=1}^{m} c_j - \sum_{j=1}^{n} d_j)\bar{Q} + \sum_{j=1}^{m} c_j (\varepsilon_{t-j} \varepsilon'_{t-j}) + \sum_{j=1}^{n} d_j Q_{t-j}.$$

In this model $\bar{Q}$ is the unconditional covariance matrix of the standardized rests $\varepsilon_t$, while $c_j$ ($j = 1, \dots, m$) and $d_j$ ($j = 1, \dots, n$) are scalar values, where $c_j$ describe the impact on current correlations of precedent shocks, and $d_j$ represent the influence on current correlations of the previous conditional correlations.

We estimate the parameters of the above copula-DCC-GARCH model using the inference function for margins (IFM) method. This method is presented in detail, among others in Joe (1997). We perform the calculations in the R environment, using the "rmgarch" package.

4. Data and results of empirical analysis.

The basis of the study are the stock quotes of 38 European insurance institutions. Most of them are on the list of the top 50 insurance companies in Europe based on total assets. AXA, a France-based company, is the largest insurance company in Europe and globally. It is also one of the world's largest asset managers with total assets under management of over 1.4 trillion euro.

Allianz, headquartered in Munich, Germany, is the second largest European insurer in terms of assets. We include insurers analyzed in the work *Network analysis of the EU insurance sector* and nine additional ones[3]. We estimate the deltaCoVaR measure assuming that the European insurance sector is represented by the STOXX 600 Europe Insurance index. We analyze weekly logarithmic rates of return for the period from January 7th, 2005 to December 20th, 2019.

In order to estimate $\lambda_t^L(i,j)$ we consider various specifications for two-dimensional copula-DCC-GARCH models. Finally, following the information criteria and model adequacy tests, we adopt for all the instruments the ARMA (1,1) -SGARCH (1,1) model with the skew Student distribution. When analyzing the dynamics of the dependences between return rates, we consider Student copulas and various DCC model specifications. As before, based on information criteria, we select the Student copula with conditional correlations obtained from the DCC(1, 1) model and a constant shape parameter. We choose the same specifications for two-dimensional copula-DCC-GARCH models, which we use to estimate the deltaCoVaR measures.

In what follows we present the results of the analysis of ten topological indicators of the MST's, divided into two groups according to their specificity. One group consists of those that are a measure of each MST vertex: Node Degree, Betweenness Centrality, Vertex Strength and Closeness Centrality. The other one is formed by those that are a measure of the properties of the entire MST, such as Average Path Length, Maximum Degree, parameters α of the power distribution of vertex degrees, Diameter, Rich Club Effect, and Assortativity.

4.1. *Degree distribution*, (Sensoy and Tabak 2014). For an undirected network, the degree $k_i$ of node $i$ is defined as the total number of links incident to it. The degree increases as a node becomes more connected and more central to the network. The degree distribution P(k) measures the frequency of nodes with different degrees in the network.

---

[3] These are: Achmea (Eureko Group), Aegon Group/Unirobe Meeùs Group, AGEAS, Allianz, Aviva, AXA, BNP Paribas, Grupo Catalana Occidente, CNP Assurances, Royal Bank of Scotland Group, Generali, Groupe Crédit Agricole Assurances, HDI/Talanx, If P&C Insurance, ING Group, KBC, Legal & General Group plc, Mapfre, Munich Re, Old Mutual plc, Prudential, RSA Insurance Group, SCOR, Lloyds Banking Group, Unipol, UNIQA Insurance Group, Vienna Insurance Group, Zurich Insurance, Swiss Life, Chubb Ltd, Hannover Re, Storebrand, XL.Group, Helvetia Holding, Mediolanum, Sampo Oyj, Societa Cattolica di Assicurazione, Topdanmark A/S.

Fig. 1. Degree distribution of selected Minimum Spanning Trees.

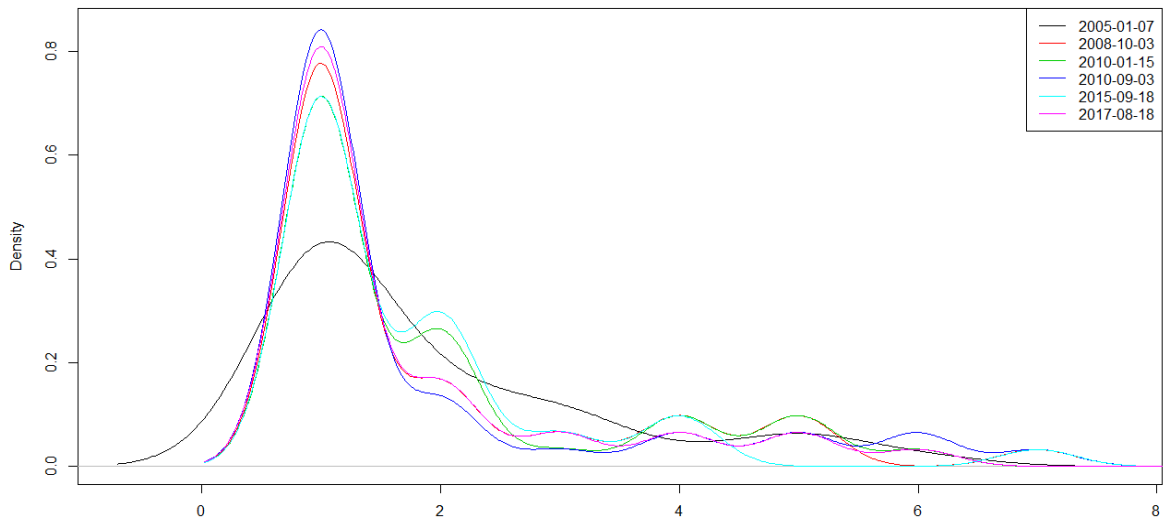

Source: *Own study.*

As depicted in Fig. 1, sample degree distribution of the minimum spanning tree is positively skewed, signalling heterogeneity in the system. Only a small portion of nodes in the network are highly interconnected (core companies), while the majority of other nodes have a relative low number of linkages (periphery companies). Such a configuration suggests the presence of several large star like structures in the minimum spanning tree. The figure highlights examples of distributions, assigned to relevant dates, which we associated during the study as outstanding. 2005-01-07 is the period preceding the subprime crisis, 2008-10-03 is the crisis, 2010-01-15 is the date of the normal state preceding the crisis of excessive public debt in the euro area, in 2010-09-03 it is a much slender distribution graph of the vertex distribution that shows the distribution during the crisis. September 18th 2015 marks the beginning of the migrant crisis in Europe. 2017-08-18 is the beginning of the crisis related to the protests in France and the "Yellow vests' movement". Therefore, periods in which we observe a high maximum value for 1 (see Fig. 1.) and at the same time low values for the remaining numbers are periods during which there are many companies with only single connections and several others having a large number of links, which is a feature favouring SR. The chart reflects the market situation.

4.2. *Betweenness Centrality – BC,* (Sensoy and Tabak, 2014). This indicator is a measure of "being between" defined as the quotient of the number of shortest paths between vertices that pass through a given vertex and the number of all the shortest paths between vertices. It determines the "most important" vertices of a given graph on a chart based on shortest paths (e.g. the most influential insurer). For each pair of vertices in a connected graph, there is at least one path between them, so that either the number of edges that you have to pass (for unweighted graphs) or the sum of the weights of the vertices that you go through (for weighted graphs) is minimized. The BC measure of a given vertex is the number of those shortest paths that pass through it. This measure defines to what extent a given node (vertex) serves as an intermediary for other network nodes. A node with a higher BC has more control over the network because more information flows through it. Figure 2 shows the mean BC for the period under consideration and each of the insurance companies studied.

Fig. 2. Average value of BC in the period under consideration for individual insurance institutions.

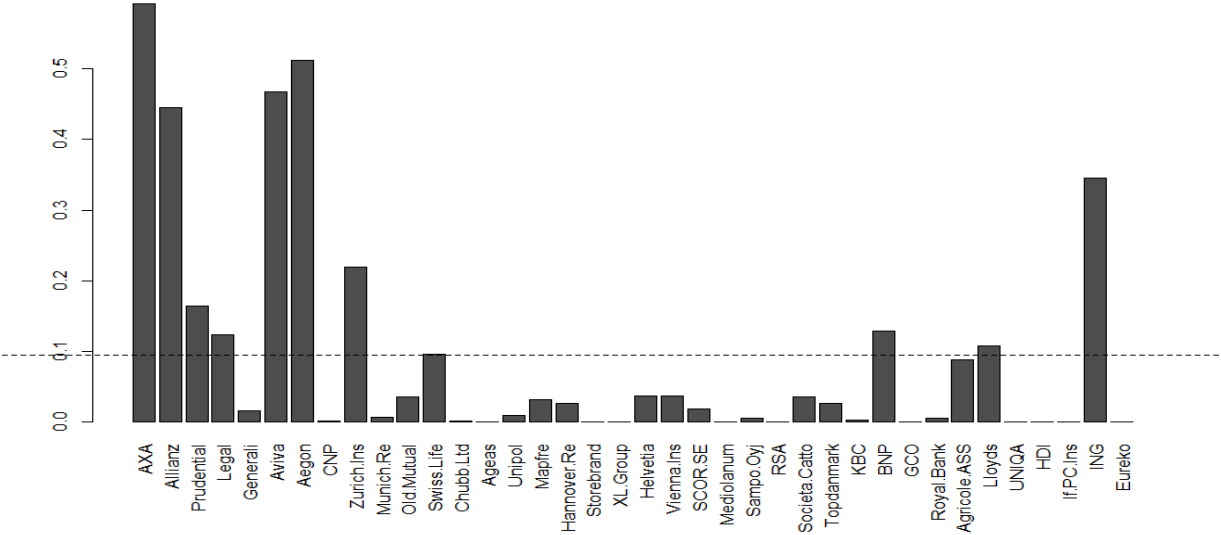

Source: *Own study.*

The analysis shows that the highest BCs is held by AXA, Aegon, Allianz, Aviva, Prudential which are companies appearing on the Financial Stability Board (FSB) 2016 list of systemically important insurers (G-SIIs), and ING, Zurich. Ins.

4.3. *Vertex strength (Centrality),* (Lautier and Raynaud, 2013). To identify the most central nodes in the system, we calculate the so-called vertex strength, which represents a weighted measure of centrality: $s_i = \sum_{j \in \psi(i)} \frac{1}{d_{ij}}$, where $\psi(i)$ is the set of all neighbors of the node $i$ and $d_{ij}$ is the length of the edge between two nodes. It indicates how far one node is from all others in the entire network. The obtained average vertex strength of the selected insurers is presented in Fig. 3.

Fig.3. Average insurers strength in the period under consideration.

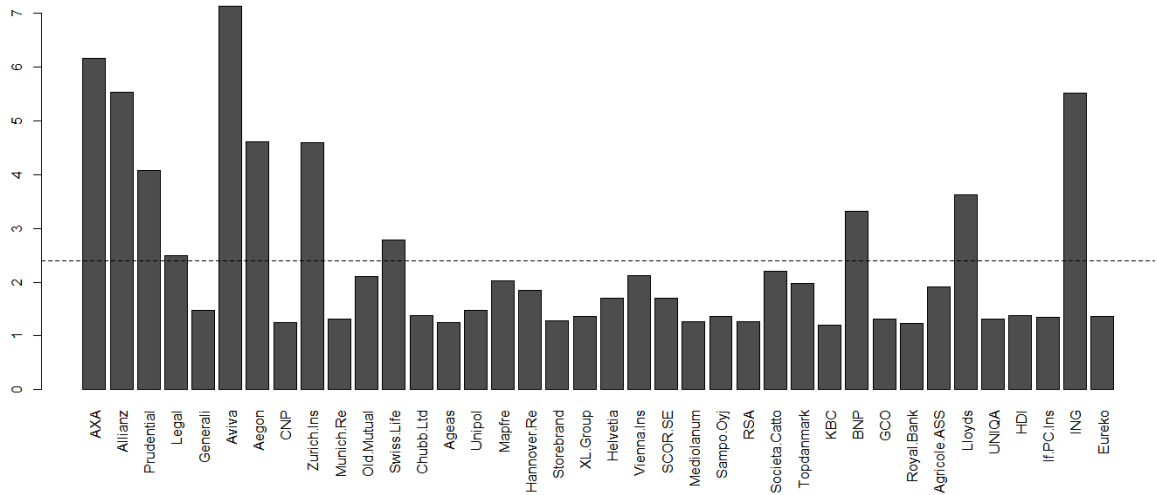

Source: *Own study*.

The higher the vertex strength, the more systemically important a node is. From the diagram above we can infer that the most important are Aviva, AXA, Allianz, ING.

4.4. *Closeness centrality*, (Bavelas 1950, Sensoy and Tabak 2014). This node proximity measure is a measure calculated as the inverse of the sum of the shortest path lengths between the given node and all nodes in the network. For MST, it is the inverse of the sum of the lengths of all edges. The more central the node is, the closer it is to all other nodes, it is thus a measure of the proximity of an insurer to the rest of the network. The average Closeness Centrality in the period studied and for each insurer considered is given in Fig. 4.

Fig.4. Average Closeness Centrality in the period under consideration.

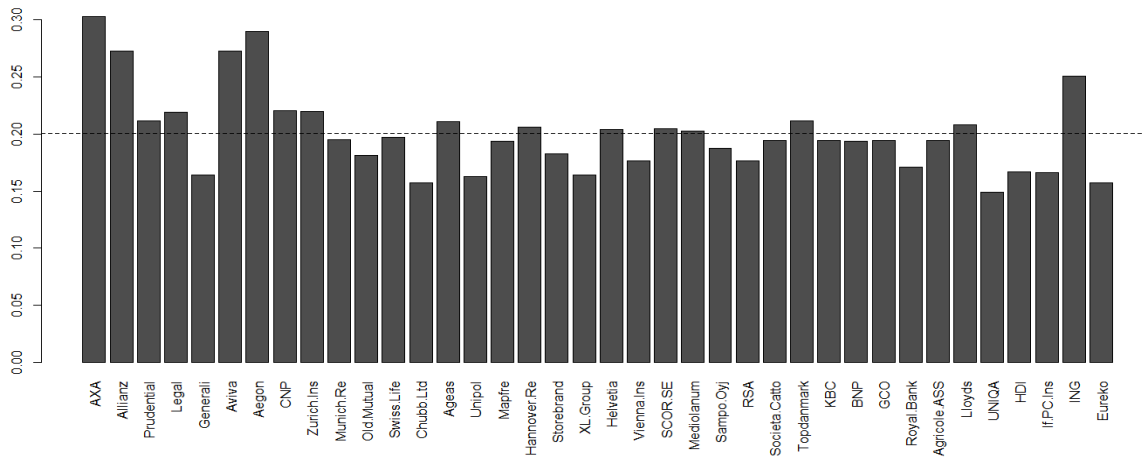

Source: *Own study.*

The diagram analysis shows that the vertices are relatively close together. This may foster contagion.

The diagram below is a summary of the four-dimensional analysis of MST indicators. We present it with the intention to draw the reader's attention to the fact that there are institutions whose bars are in each case considered among the highest ones, which proves the importance of their corresponding vertex in the MST.

Fig. 5. Average Betweenness Centrality, Vertex degree, Vertex strength and Closeness Centrality.

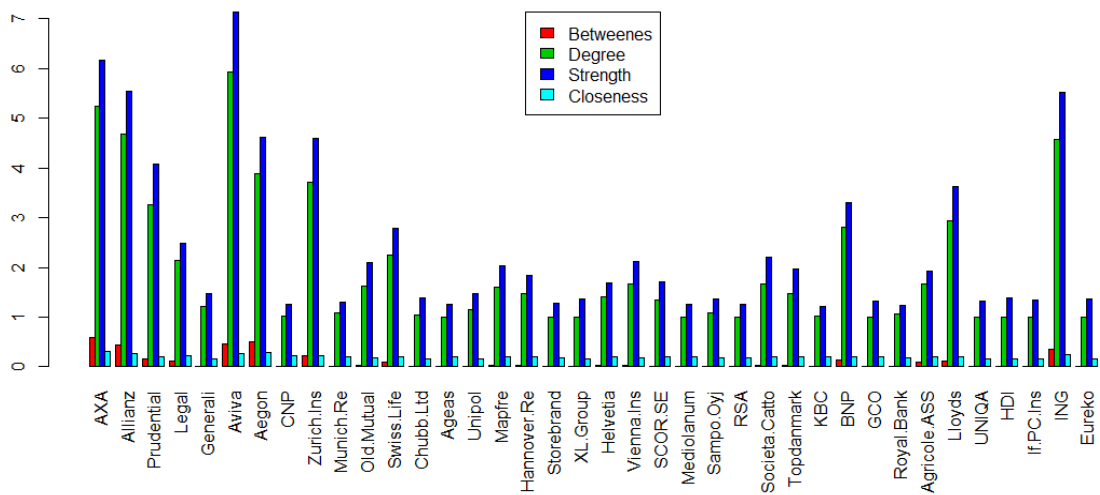

Source: *Own study*

4.5. *Average Path Lentgth (APL),* (Wang et al., 2014) This indicator is defined as the average number of steps along the shortest paths for all possible pairs of network nodes. It measures the effectiveness of information flow or mass transport in a given network. APL is one of the strongest measures of network topology, along with its clustering factor and degree distribution. It distinguishes an easy-to-access network from a more complex and inefficient one. The smaller the average path length, the easier the flow of information. Of course, we are talking about average so the network itself can have several very distant nodes and many adjacent nodes. The times series obtained for the APL is presented in Figure 6.

Fig. 6. Average Path Lentgth in the period 07.01.2005-20.12.2019.

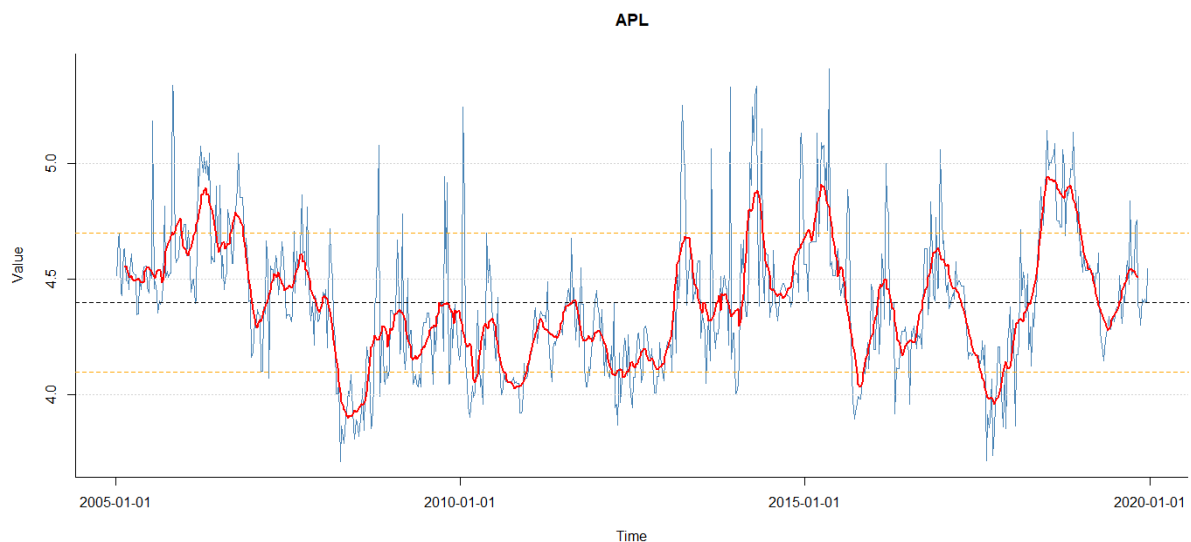

*Red lines depict the 13 periods moving average smoothed series.
Source: *Own study*

Note that during the crisis, the APL indicator decreases in comparison to the normal state, which means that the average path length between any pair of companies decreases.

4.6. *Maximum Degree - Max.Deg,* (Wang et al., 2014). This indicator in graph theory is the maximum degree (i.e. the number of edges coming out of it, where loops count double) of a vertex of the graph; it is thus the number of connections of a central vertex. The times series obtained for the Maximum Degree is shown on Figure 7. This indicator in graph theory is the maximal number of connections a vertex of the graph has. By connections we mean, of course, the number of edges coming out of a vertex, with loops counting double.

Fig.7. Maximum Degree in the period 07.01.2005-20.12.2019

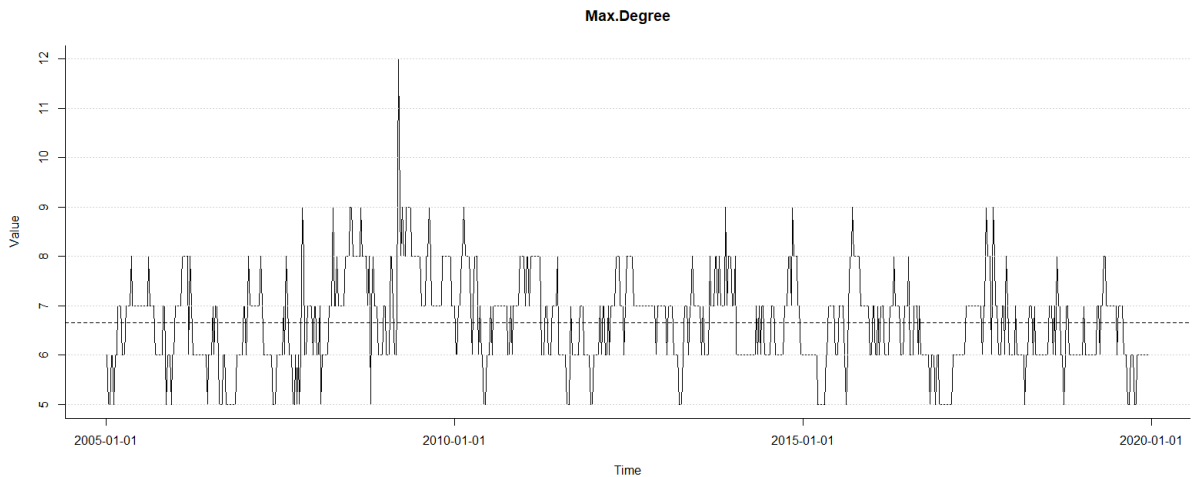

Source: *Own study.*

Maximum Degree grows during periods of crisis, which means that in a group of insurers during a crisis some insurer has many more connections with others than is usual in the normal state.

4.7. *Parameter α of the vertex degree distribution* (Wang et al., 2014) required to follow a power law. This indicator measures the scale-free behavior of a network. The network is scale-free if the distribution $P(s)$ the number of connections between the vertices is follows a power law, i.e. it has (asymptotically) the form $P(s) = C \cdot s^{-\alpha}$, $\alpha > 0$, where $\alpha$ is a parameter specific to the given network. The power law followed by the degree distribution gives the network a kind of fractal self-similarity properties, which accounts for the name. A scale-free network is characterized by a small number of vertices having a large number of connections (such nodes are called hubs) and many vertices that have only one connection. From the point of view of our analysis, this type of network can be considered as "favorable" to the propagation of information (in our case: of systemic risk), and the companies-hubs that it has are systemically relevant. The time series obtained for the alpha parameters is shown on Figure 8.

Fig.8. Estimated parameters alpha of power distribution for the MST from 07.01.2005 to 20.12.2019

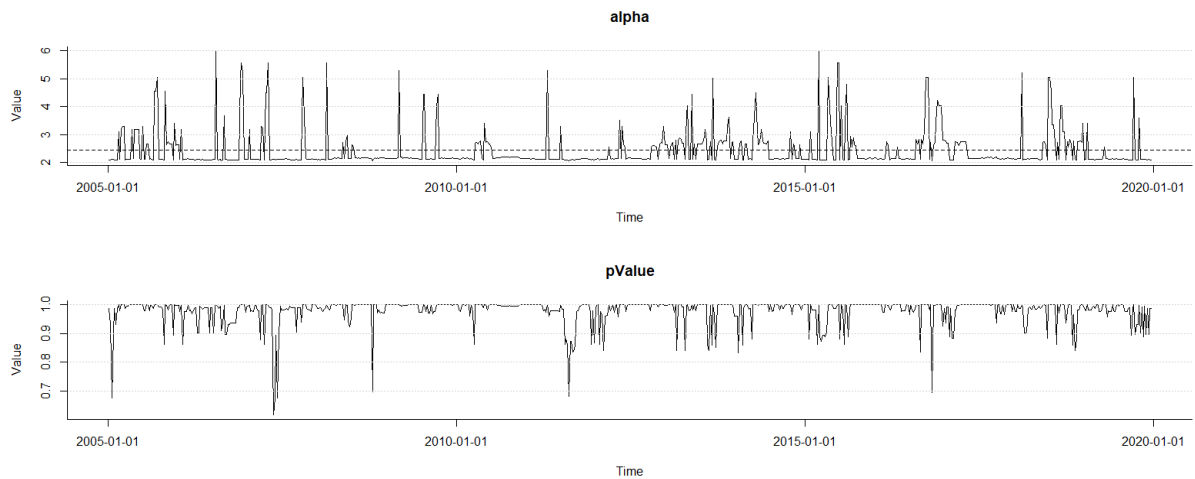

Source: *Own study.*

MSTs are scale-free, but during crises the alpha value is closer to 2, which means that the structure of MST is star-shaped with outstanding hubs having a high degree, i.e. multiple edges that connect the company-hub to several companies with only one edge.

4.8. *Diameter of the network* (Diameter). It is determined by choosing from among all the shortest paths connecting any pair vertices the longest one. For MST, this is simply the longest path in the MST. The time series obtained for the Diameter is shown in Figure 9.

Fig. 9. Diameter in the period 07.01.2005-20.12.2019

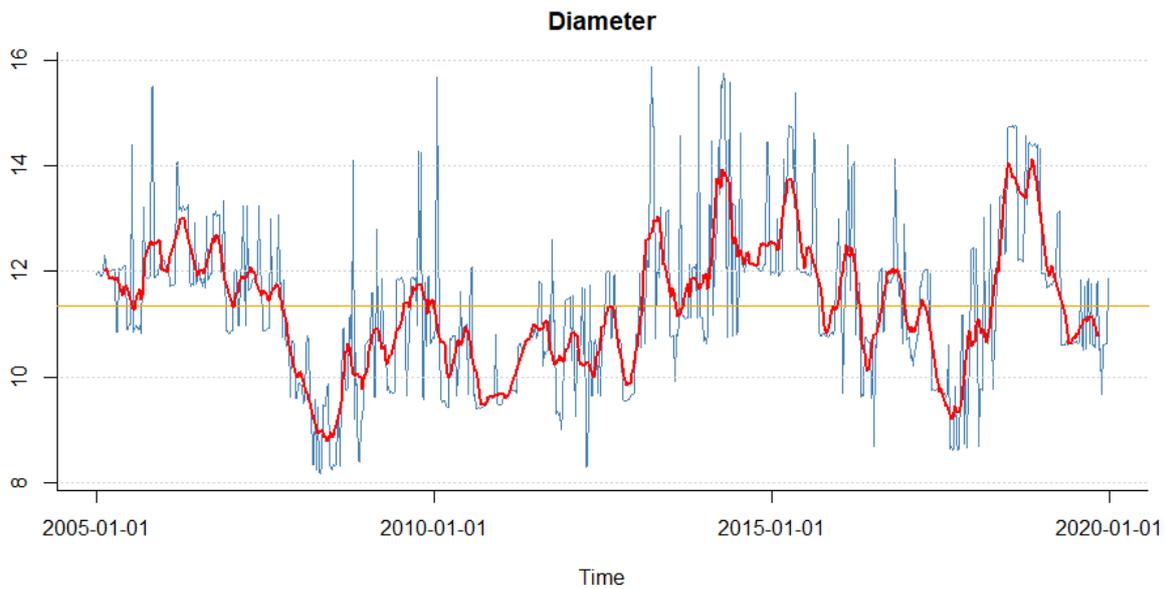

*Red lines depict the 13 periods moving average smoothed series.
Source: *Own study*

Diameter decreases during crises, which means that during these periods the path between the further apart lying MST vertices is shortened.

*4.9. Rich Club Effect – RCE,* (Colizza et al., 2006) The idea is that well-connected vertices connect also one with another. The RCE is defined to be $\phi(k) = \frac{2E_{>k}}{N_{>k}(N_{>k}-1)}$, where $\frac{N_{>k}(N_{>k}-1)}{2}$ is the number of all the possible paths between $N_{>k}$ vertices, $E_{>k}$ is the number of vertices of $N_{>k}$ nodes having degree $>k$. The effect of a rich club reduces system stability, which means that if RCE increases, then a perturbation can be more easily transmitted through the network. The times series RCE for k=4 obtained in the study is shown in Fig. 10. while its distribution in the market states determined is presented in Fig. 13. They show the results for the dynamic MST with ϕ (4).

Fig. 10. RCE during the period 07.01.2005-20.12.2019

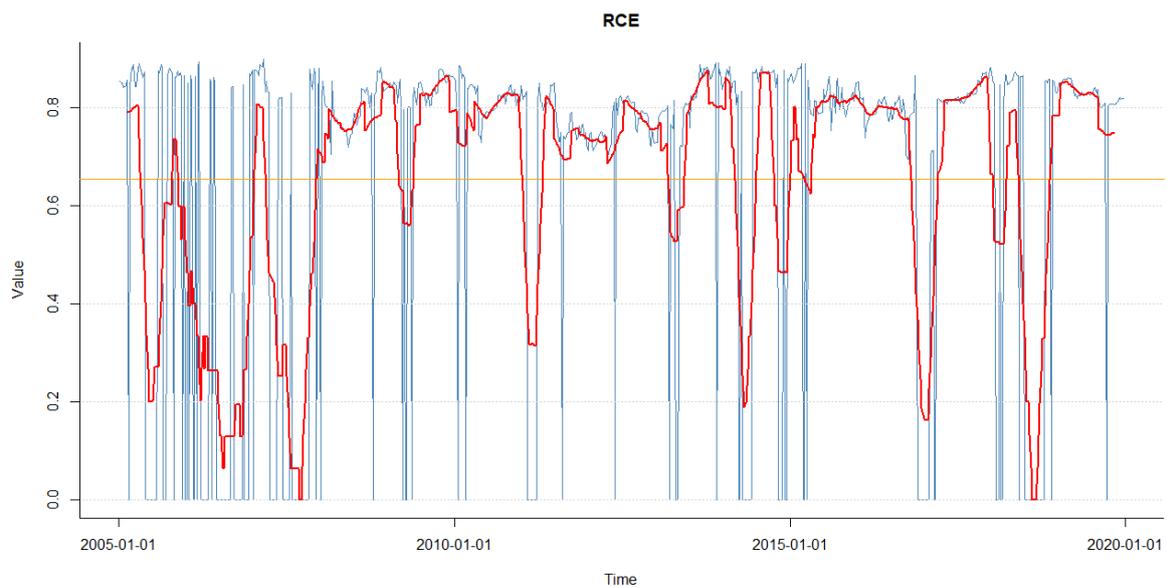

*Red lines depict the 13 periods moving average smoothed series.
Source: *Own study*

4.10. *Assortativity.* The concept of assortativity was introduced by Newman (2002) and has been intensively studied since then. Assortativity is a graphic measure. It shows to what extent nodes in the network associate one to another by similarity or opposition (positive or negative mating). Basically, the network's assortatavity is determined for the degree (number of direct neighbors) of nodes in the network. Assortativity is expressed as a scalar $-1 \leq \rho \leq 1$. The network is said to be assortative when high-degree nodes are mostly connected to other high-degree nodes while low-degree nodes are mostly connected to other low-degree nodes. The network is said to be non-assortatative when high-degree nodes are connected mostly to low-degree nodes and low-degree nodes are mostly connected to high-degree nodes. Assortativity provides information on the structure of the network, but also on its dynamic behavior and robustness. The assortativity time series is shown in Fig. 11.

Fig. 11. Assortativity during the period 07.01.2005-20.12.2019.

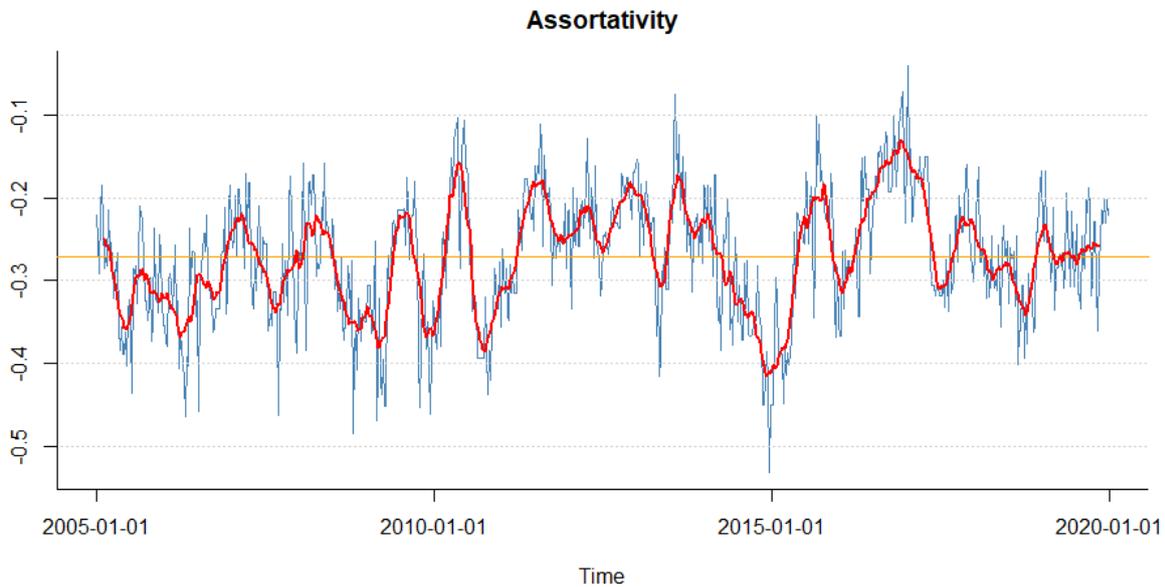

*Red lines depict the 13 periods moving average smoothed series

Source: *Own study.*

Assortativity is negative, which means that in each state the tree is rather non-assortatative, i.e. the vertices tend to connect rather as negative mating. Which also confirms the previously described property of the network to be scale-free.

Based on the time series of MST topological indicators, we determined a time partition into four periods:
- the period which we call normal state – Normal (N)
- a period of two subprime crises and excessive public debt, which began in 2008 and lasted until around 2013. This period in our time series falls exactly between February 8th, 2008 and March 1st, 2013 – Subprime Mortgage Crisis (SMC)
- the period of crisis associated with the beginning of the migration crisis in Europe, falling on 2015 / 2016. This period on our time series falls exactly between August 7th, 2015 and on September 23rd, 2016 – Immigrant (I)
- the period of the beginning of the crisis in the countries of the European Union related to the crisis in France associated with strikes, and in Italy due to the ever-growing public debt (which is now seven times higher than the debt in Greece), falling at the turn of 2017 and 2018. In our case it is exactly the period from April 21st, 2017 until May 11th, 2018 – (FIC)

The charts below (Fig. 12.) present the expected values of the relevant MST topological indicators in the different market states. The results confirm the above description of indicators during crises and in normal state. Putting together the indicators allows us to compare their behaviour in four different states. Clearly, the indicators behave differently during crises. In state N, APL and Diameter is higher, while Maximum Degree is lower than in crisis states. This means MST is stretching. The insurer with the largest number of connections has actually fewer connections than during the crises. RCE varies, but the average - marked with a red dot - is smaller, i.e. the stability of the network is higher. The network assortativity in each state is at a similar, negative level, i.e. the network does not vary, remaining constantly similar. Connections are established by negative mating. High degree companies are linked to low degree ones. Because alpha is close to 2, MSTs are scale-free, but the average alpha increases during the period N.

Fig. 12. Distribution of MST topological indicators in different market states.

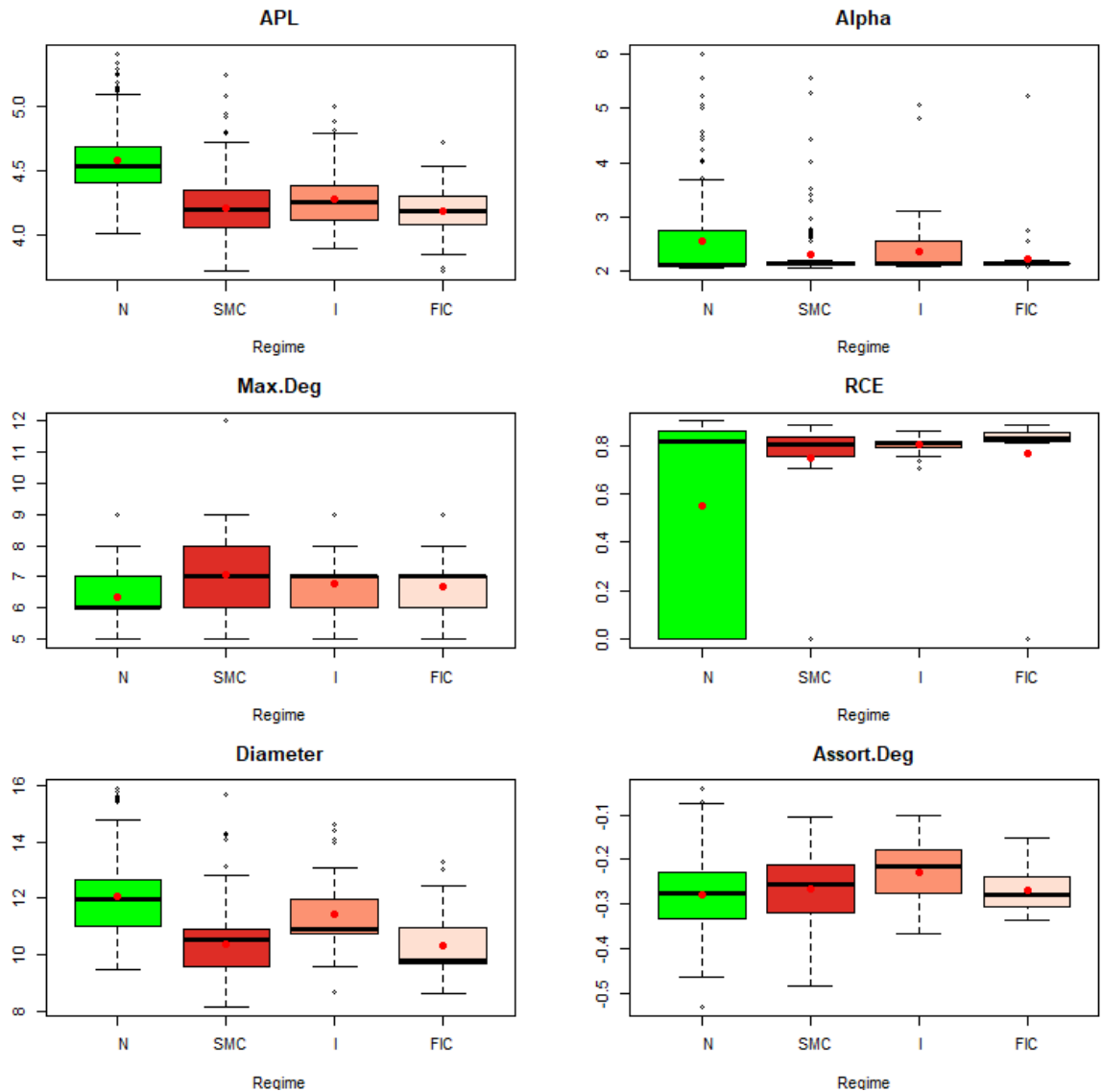

Source: *Own study.*

Special attention is drawn to the RCE indicator (Fig. 12.), whose average value in the state N is very different from the average in the remaining distinguished states. By analyzing in more detail (Fig. 13.), we note that in the states presented RCE is high, which means that the way the vertices are connected one to another is such that the highest degree vertices are linked together. Potentially, this creates the possibility of transferring turbulences. In the normal states the series shows an important variability and the average RCE is high. In the state N the mean is lower. RCE takes the value zero many times. This means that there are no vertices with four or more edges. MST appears in the form of a stretched chain. It is then more stable than during the SMC, I or FIC periods.

Fig. 13. RCE distribution in determined market states based on the mean and the standard deviation for k=4.

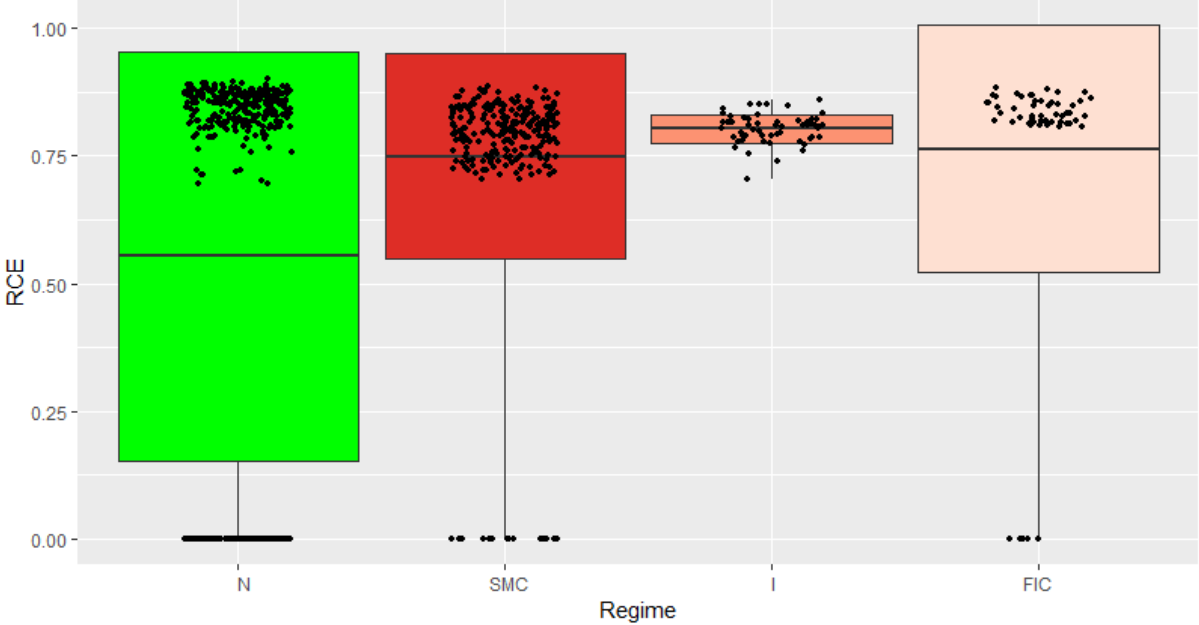

Source: *Own study.*

Below (Fig. 14) are sample MSTs at selected times: a tree that has a chain-like structure will slow down risk propagation, and one that has a star-shaped structure will foster it.

Fig. 14. Sample MSTs in two chosen times.

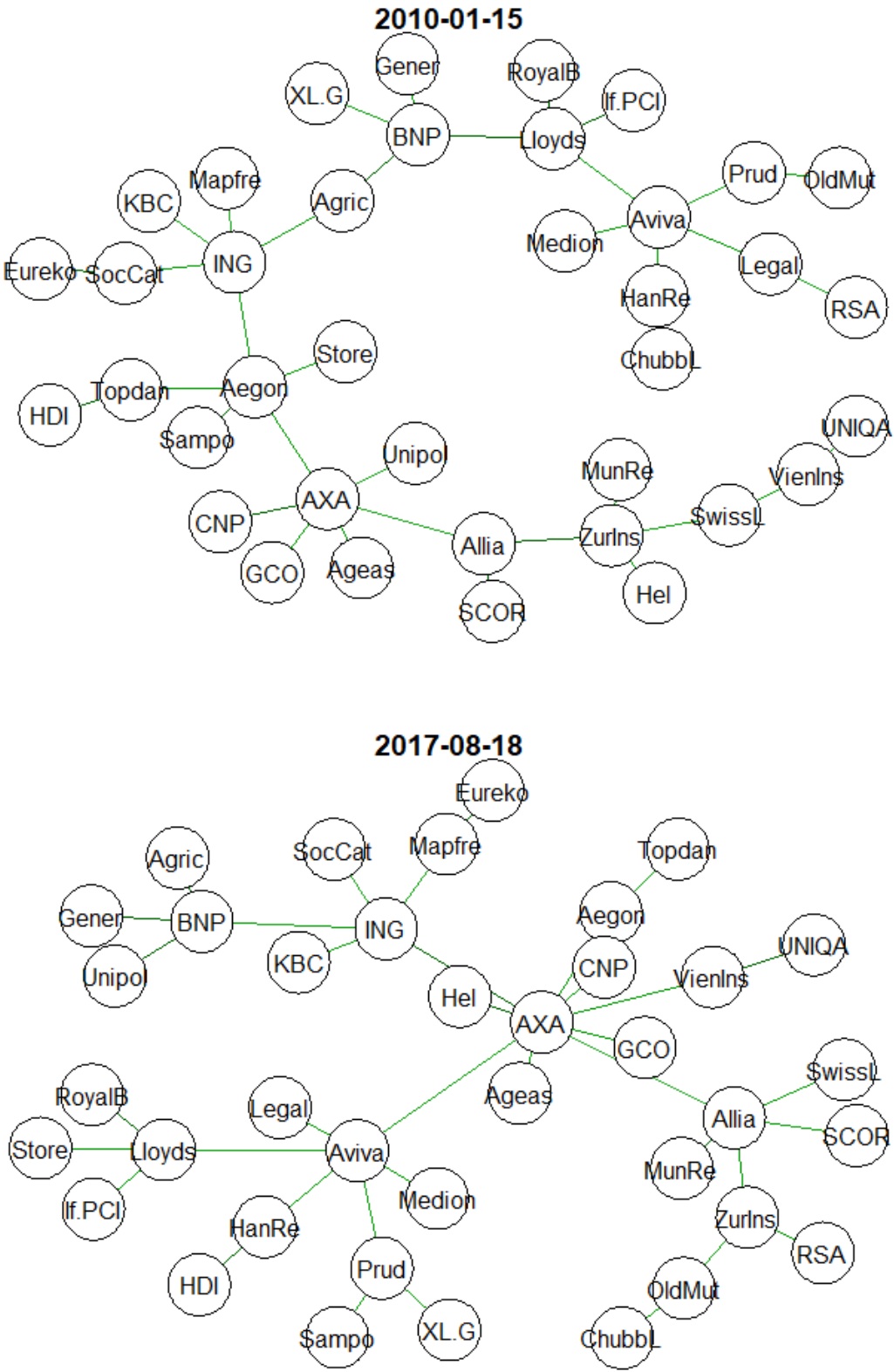

Source: *Own study.*

On figures 15, 16, 17 we present the results obtained for deltaCoVaR. These are respectively the average deltaCoVaR value in the period studied, the distribution of this average in the different market states and the average deltaCoVaR value for each insurer in the period under consideration. They corroborate the fact that insurance companies contribute to systemic risk. This contribution depends on the market state.

Fig. 15. Mean deltaCoVaR.

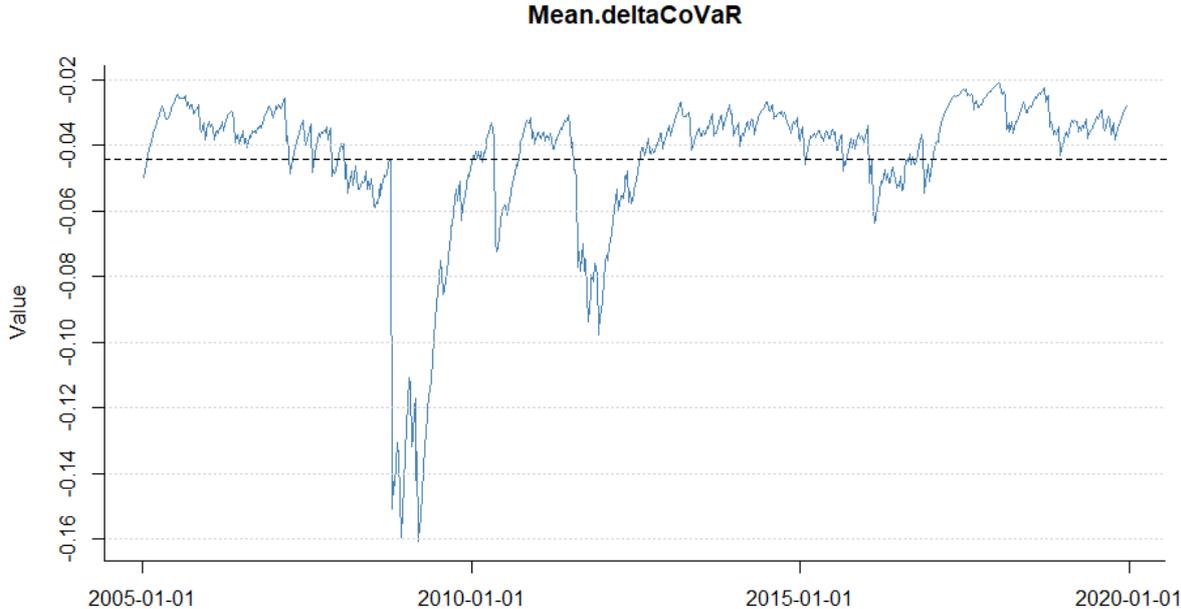

Source: *Own study*.

The Mean.deltaCoVaR chart confirms the fact that insurance companies contribute to systemic risk. deltaCoVaR decreases in highlighted periods of crises.
On average, each company contributes to systemic risk.

Fig.16. Mean deltaCoVaR for each insurer.

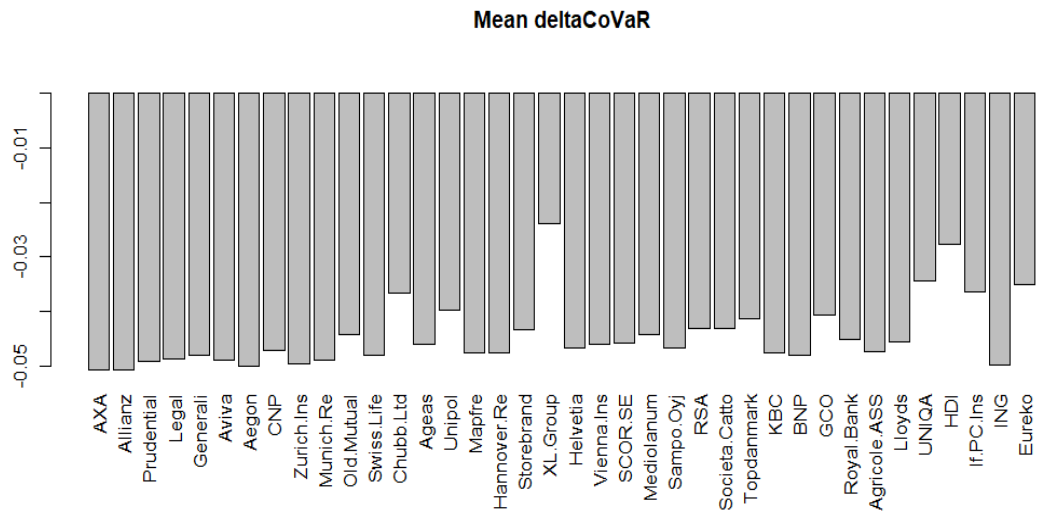

Source: *Own study.*

Fig. 17. Mean deltaCoVaR distribution in different market states.

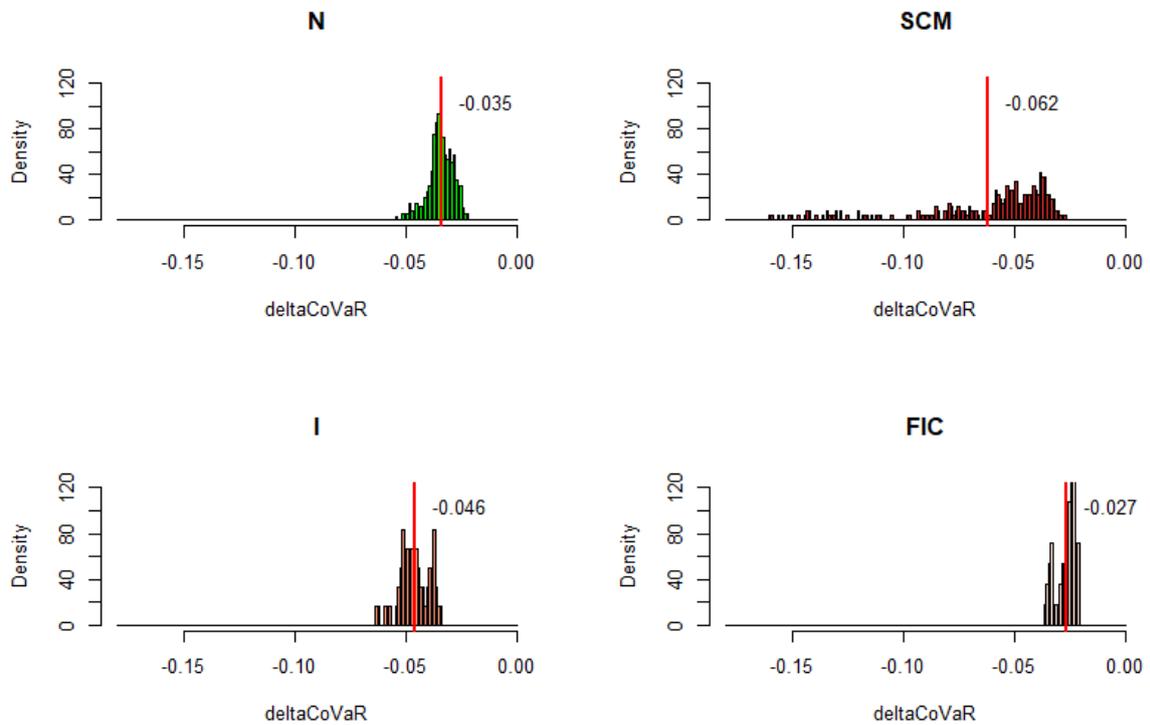

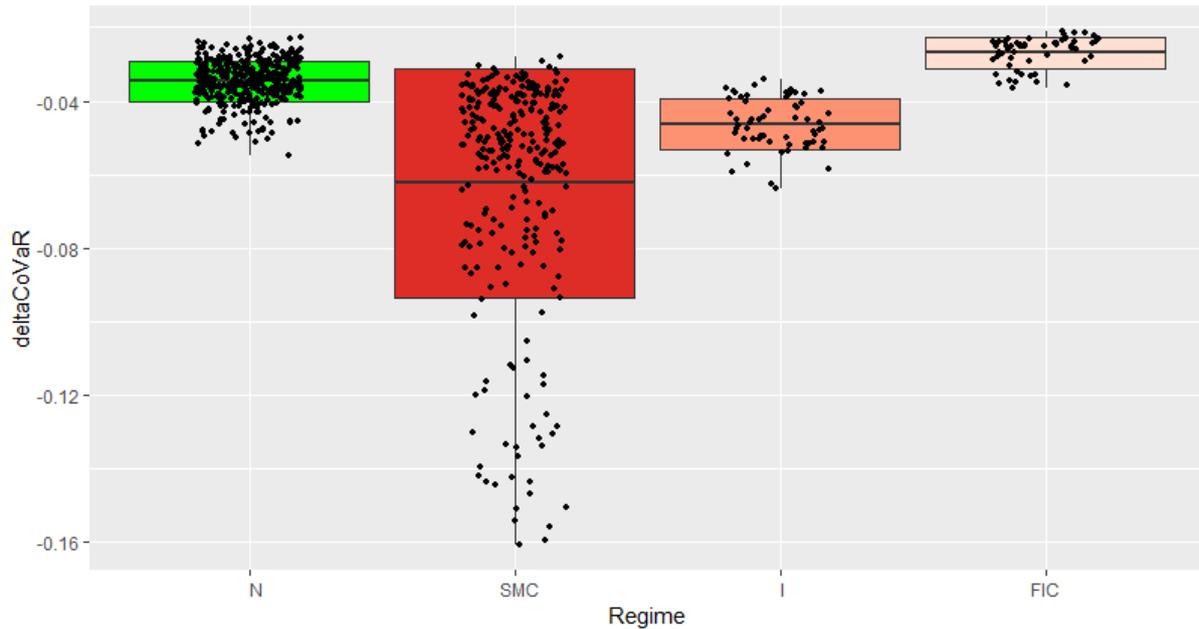

Source: *Own study.*

The smallest contribution to SR is observed in the FIC and N periods. The largest contribution in the SMC period. The beginning of immigration was also a period in which we notice an increased contribution to the SR.

5. Conclusions.

Empirical results show the usefulness of network topology indicators for detecting systemic risk in the insurance sector. The proposed hybrid approach can be used to predict systemic risk in the insurance sector. Analysis of time series also including the most current data from the period until the end of 2019 confirms the market phenomena. In the distinguished states, the structure of MST changes along with the market situation. MSTs – their topological indicators are a tool that also allows clustering in the insurance sector and helps determine those companies that are relevant in the entire group. AXA, Allianz, Aegon, Aviva and non-G-SIIs ING and Zurich Insurance play a significant role. These are institutions through which risk may be transferred. They stand out when assessing BC, Strength, Closeness Centrality, Degree. In assessing the indicators for the entire network, we note that in the three periods of crises we have identified, APL and Diameter decrease, Maximum Degree increases, RCE is high, the assortment is negative. All this means that in the SMC, I, FIC states MST becomes star-like and compact, which is accompanied by a decrease in system stability, which means that turbulence can be more easily transmitted over the network. Such a network configuration is less resistant to shocks and more susceptible to contagion and transferring the effects of collapses on the financial market. All institutions contribute to SR, and again the largest contribution is due to the previously mentioned companies. When analyzing deltaCoVaR averages, the crisis period again stands out. In the so-called period N, the contribution is the smallest.

**References.**